\pdfoutput=1

\documentclass[twocolumn,showpacs,amssymb,aps,nofootinbib,floatfix,groupedaddress]{revtex4-1}

\usepackage{epsfig}
\usepackage{hyperref}
\usepackage{comment}
\usepackage{amsmath}
\usepackage{graphicx,color}

\begin{document} 
\hbadness=10000

\title{Effect of bulk viscosity  on interferometry correlations in
ultrarelativistic heavy-ion collisions}

\author{Piotr Bo\.zek}
\email{piotr.bozek@fis.agh.edu.pl}
\affiliation{AGH University of Science and Technology, Faculty of Physics and
Applied Computer Science, al. Mickiewicza 30, 30-059 Krakow, Poland}

\begin{abstract}
A temperature dependent bulk viscosity coefficient is used in $3+1$-dimensional 
hydrodynamic simulations. I study the effect of the increase of bulk 
viscosity around the critical temperature on the system dynamics in
central Pb+Pb collisions at $\sqrt{s_{NN}}=2760$~GeV. With
 increasing bulk viscosity the life-time of the system increases slightly.
Also the shape of the freeze-out hypersurface  changes, the outer layers of the fireball live longer. This effect causes a small reduction of the ratio of two interferometry radii $R_{out}/R_{side}$, improving the agreement with experimental data.  
\end{abstract}

\date{\today}



\maketitle


\section{Introduction \label{sec:intro}}
The interferometry correlations are a
sensitive measure of the space-time evolution  of the system formed in heavy-ion collisions. The interferometry, or Handburry-Brown Twiss (HBT) radii, extracted from the same-sign $\pi$-$\pi$ correlation function give an estimate of the  size of the region where the pion pair is emitted
\cite{Lisa:2005dd,Wiedemann:1999qn}. From the studies of Au-Au collisions at BNL  Relativistic Heavy Ion Collider 
 energies it has been found that hydrodynamic simulations describe fairly well the experimental data on HBT correlations 
\cite{Broniowski:2008vp,Pratt:2008qv}. Similar agreement is seen for Pb+Pb collisions at the CERN Large Hadron Collider.

From the analysis of the three dimensional $\pi$-$\pi$ correlation function three radii parameters can be extracted
 \cite{Bertsch:1989vn,Pratt:1986cc}, $R_{out}$, $R_{side}$, and $R_{long}$.
A key assumption required to reach agreement with the data  is the use of a hard equation of 
state for the matter created in the collision. A softening of the equation of state  leads 
to an increase of the $R_{out}/R_{side}$ ratio \cite{Rischke:1996em,Bozek:2009ty}, that would spoil the agreement with the experimental
 data. 
The measurement  of the  $R_{out}/R_{side}$ value 
close to one suggests a rapid expansion of the fireball. Besides a hard equation of state, 
other effects that enable transverse expansion lead to a reduction of $R_{out}/R_{side}$. The factors that influence the 
transverse expansion  are shear viscosity, 
preequilibrium flow or steeper gradients in the initial entropy distribution \cite{Broniowski:2008vp,Pratt:2008qv,Bozek:2010er,Karpenko:2012yf}.
Using a hard equation of state and realistic assumptions for the initial density 
profile and the preequilibrium flow the observed ratio $R_{out}/R_{side}$ can be reproduced in hydrodynamic models to within $10\%$.

A sizable value of the bulk viscosity may slow down the expansion of the fireball.
It is expected that around the QCD critical temperature the value of  bulk viscosity rises
\cite{Karsch:2007jc}. This peak in the temperature dependence of  bulk viscosity  has been implemented in a number of
hydrodynamic simulations 
\cite{Denicol:2009am,Song:2009rh,Ryu:2015vwa,Bernhard:2016tnd}. The conclusion from these studies is that 
bulk viscosity of matter influences the
 transverse expansion
of the fireball. Bulk viscosity leads to a reduction of the effective pressure.

This rises the question, whether introducing a significant bulk
 viscosity around the critical temperature would not invalidate the good description 
of the interferometry data. In an expanding system 
 bulk viscosity reduces the effective pressure and acts in a similar way
 as a soft point in 
the equation of state. In this paper I investigate this question  in explicit hydrodynamic simulations for central Pb-Pb collisions.
Contrary to naive expectations, I find that the presence of a large bulk viscosity peak around the critical temperature does not lead to an increase of the $R_{out}/R_{side}$ ratio. In fact one  
 finds  a small decrease in that ratio for simulations with bulk viscosity, 
driving it closer to the experimental value.

\section{Model and calculations}

\begin{figure}[tb]
\includegraphics[width=0.3 \textwidth]{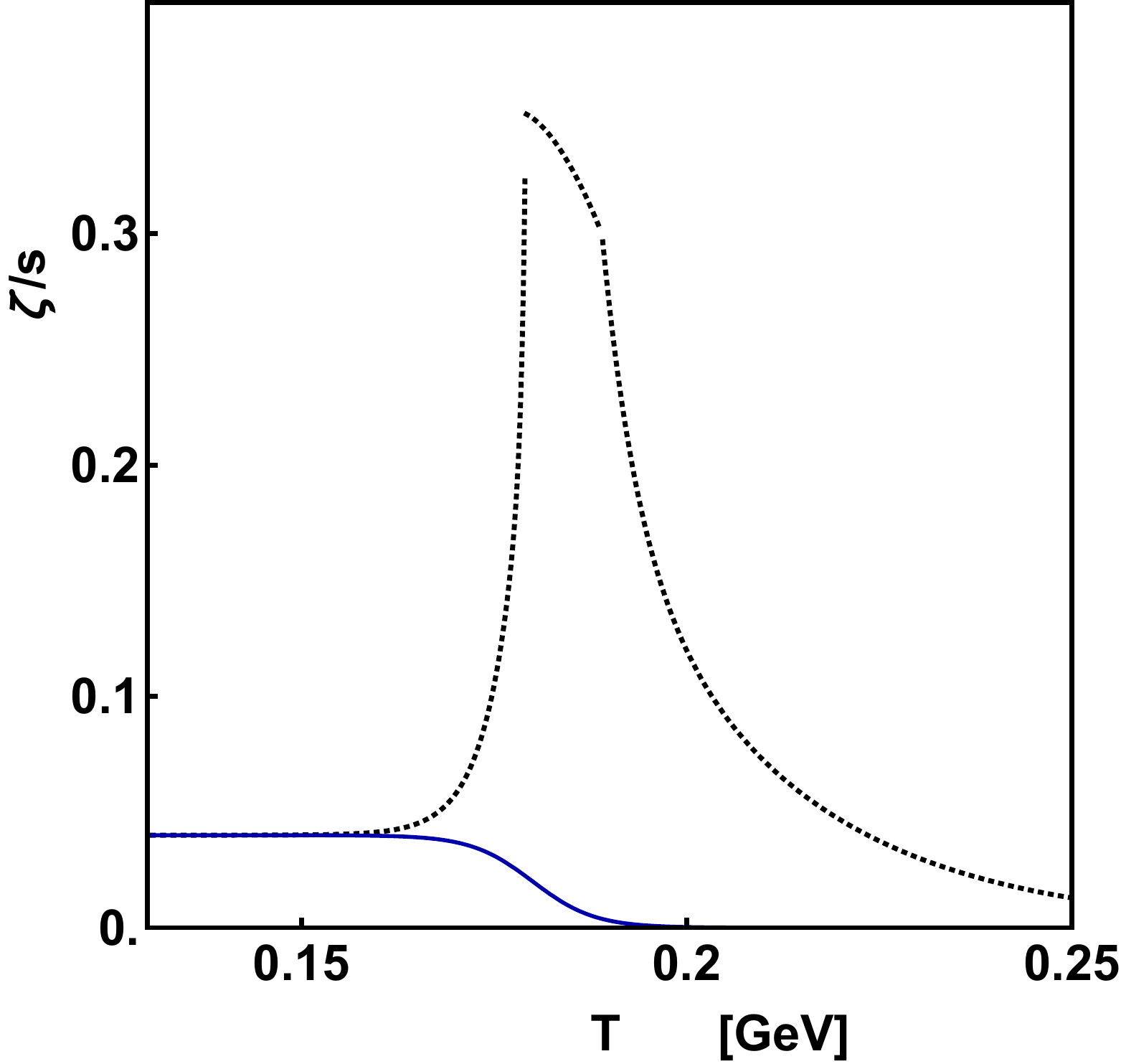}
\caption{(color online) The temperature dependence of the ratio of the bulk viscosity coefficient to entropy density $\zeta/s$ 
for the two scenarios considered, with bulk viscosity in the low temperature only (solid line) and  with an additional
 peak of bulk viscosity
around the critical temperature (dotted line).
\label{fig:zeta}}
\end{figure}

I perform calculations in a $3+1$-dimensional viscous hydrodynamic model
 \cite{Schenke:2010rr}. The shear viscosity to entropy density ration 
is $\eta/s=0.08$. The bulk viscosity to entropy ratio is temperature dependent
\begin{equation}
\zeta/s = \frac{\zeta_{H}}{s} \frac{1}{1+\exp\left( \left(T-T_H \right)/
\Delta T_1\right)}+ c \frac{\zeta_{peak}}{s}\left(T \right) 
\end{equation}
where
\begin{equation}
\frac{\zeta_{peak}}{s}\left(T\right)=  \begin{cases}
l_1  \exp\left(\frac{T/T_{H}-1}{s_1}\right)+l_2  \exp\left(\frac{T/T_H-1}{s_2}\right)
 & \\   \text{for} \ \ T<T_a \\
A_0+A_1\frac{T}{T_H}+A_2 \left(\frac{T}{T_H} \right)^2
 & \\   \text{for} \ \ T_a<T<T_b \\
l_3  \exp\left(\frac{1-T/T_{H}}{s_3}\right)+l_4  \exp\left(\frac{1-T/T_H}{s_4}\right)
 & \\   \text{for} \ \ T>T_b 
\end{cases}
\end{equation}
and $T_H=180$MeV, $T_a=179.1$MeV, $T_ b=189$MeV, $\Delta T_H=4$MeV,  $A_0 = -13.45$, $A_1 = 27.55$, $A_2 = -13.77$, $l_1 = 0.9$, $l_2=0.26$, 
 $l_3=0.9$, $l_4=0.256$, 
$s_1 = 0.0025$, $s_2 = 0.022$, $s_3 = 0.025$, $s_4 = 0.13$.
The parametrization of the bulk viscosity peak is similar as in 
\cite{Denicol:2009am}.
 In the following I study two different cases $c=0$,  $1$ (Fig. \ref{fig:zeta}).  The first case corresponds
 to no bulk viscosity peak, with bulk viscosity appearing only at low temperatures. Such a small bulk viscosity appears 
naturally in a mixture of massive particles with nonequilibrium distributions.
 The parametrization with $c=1$ includes a bulk viscosity 
peak around the critical temperature
as expected from lattice QCD results \cite{Karsch:2007jc}. The height of the bulk viscosity peak is also estimated 
from fits to particle spectra and azimuthal harmonic flow coefficients \cite{Bernhard:2016tnd}.
In the following I denote the two scenarios by  the value of  bulk viscosity at $T_H$, that  is $\zeta/s=0.02, \ 0.35$. 
An increase 
 in the bulk viscosity implies larger dissipative correction in the hydrodynamic evolution. The corresponding 
entropy production in the evolution is compensated by a reduction of the initial entropy for calculation with
 larger bulk viscosity to obtain the same final particle multiplicity at central rapidity.

The results presented further correspond to the  freeze-out temperature of $150$~MeV. The Cooper-Frye formula 
for the emission of hadrons from the freeze-out hypersurface includes nonequilibrium corrections from both shear 
tensor and bulk viscosity. Bulk viscosity corrections are included using the relaxation time approximation \cite{Bozek:2009dw}. 
I have checked 
that varying the freeze-out temperature down to $140$~MeV does not change the conclusions. 

From  pion pairs  a  Bose-Einstein symmetrized correlation function is constructed \cite{Chojnacki:2011hb} and
binned in bins of average transverse momentum of the pion pair. The three dimensional correlation function in relative pair momentum $q$ 
is
fitted with a Gaussian formula  \cite{Bertsch:1989vn,Pratt:1986cc}
\begin{equation}
C(q)=1+\lambda \exp\left(-R_{out}^2q_{out}^2-R_{side}^2q_{side}^2-R_{long}^2q_{long}^2\right) \ .
\label{eq:pb}
\end{equation}
The three components of $q$ are defined along the beam axis ($q_{long}$), along the pair transverse momentum ($q_{out}$), and
transverse to those ($q_{side}$). The  interferometry analysis is summarized by the three HBT radii $R_i$ as functions
 of the average pair momentum $k_T$.

\section{Smooth initial conditions}

To present a simple example I calculate the evolution of the fireball with smooth initial density density. The entropy density is
 obtained from the optical Glauber model. The  density in the transverse plane is proportional to a combination of 
densities of participant nucleons $\rho_w(x,y)$ and binary collisions $\rho_{bin}(x,y) $
\begin{equation}
s(x,y) \propto (1-\alpha) \rho_W(x,,y) + 2 \alpha \rho_{bin}(x,y) \ ,
\end{equation}
with $\alpha=0.15$ \cite{Bozek:2011ua}. The normalization of the initial density is adjusted 
in order to reproduce the  charged particle multiplicity. The calculation is performed for zero impact parameter for illustration.

\begin{figure}[tb]
\includegraphics[width=0.3 \textwidth]{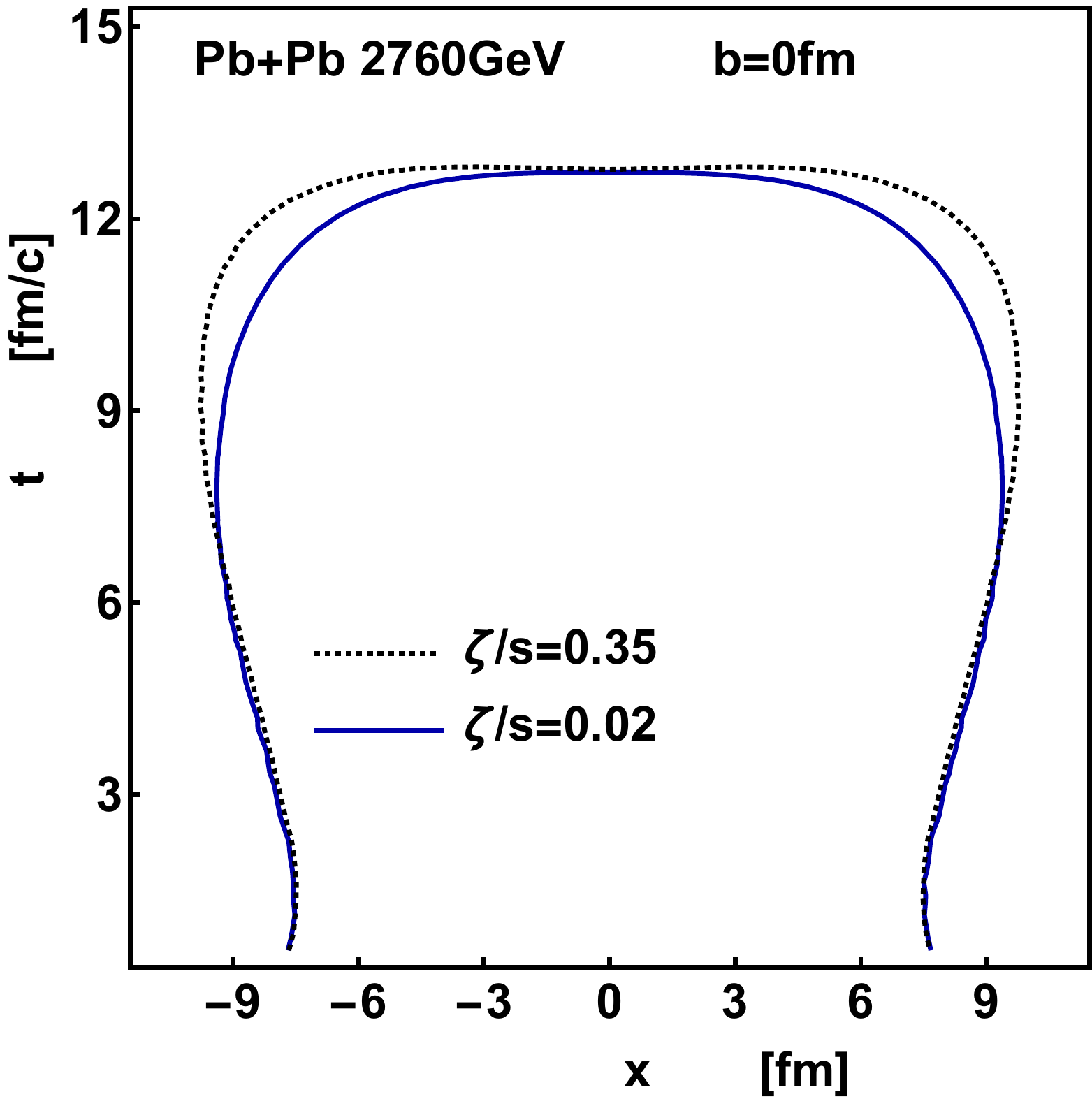}
\caption{(color online) The shape of the freeze-out hypersurface from
 hydrodynamic simulations with two scenarios for the bulk viscosity coefficient, with and without a bulk viscosity peak (dotted and 
solid lines respectively). 
\label{fig:fr1}}
\end{figure}

The bulk viscosity peak reduces the transverse flow at the edge of the fireball. It causes a
 retardation of the expansion at the edge. The resulting freeze-out hypersurface is slightly modified (Fig. \ref{fig:fr1}). 
The effect on the freeze-out hypersurface can be described as a change from a scenario with earlier freeze-out in the outer 
layers (``burning log scenario'') to 
a freeze-out at constant proper time. The
total life-time of the system is similar in the two scenarios.

\begin{figure}[tb]
\includegraphics[width=0.52 \textwidth]{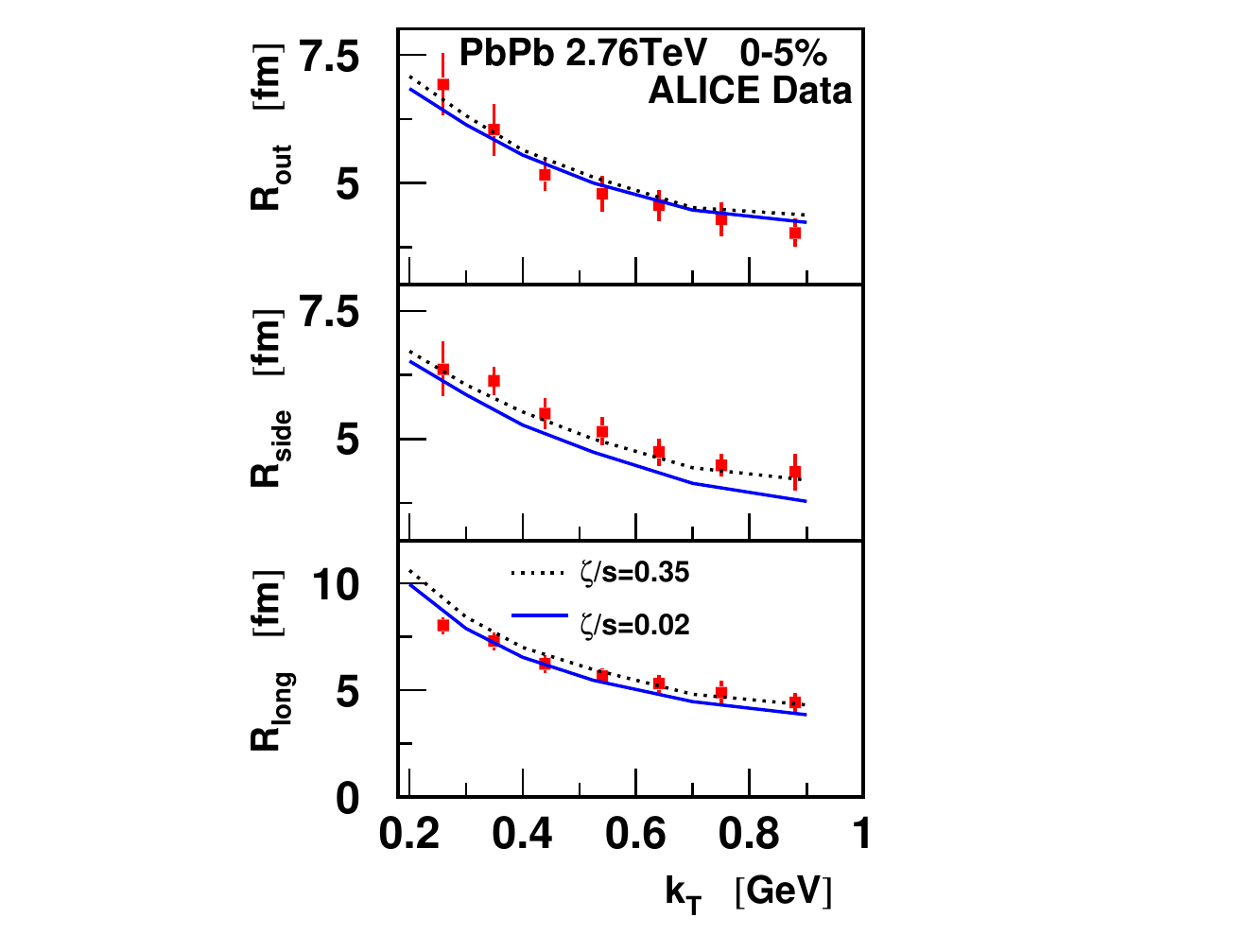}
\caption{(color online) The HBT radii for central Pb-Pb collisions at 2760GeV. ALICE Collaboration data (squares) compared to hydrodynamic calculations with smooth initial conditions, with (dotted lines) and without (solid lines) a peak in the temperature dependence of bulk viscosity.
\label{fig:hbt1}}
\end{figure}

\begin{figure}[tb]
\includegraphics[width=0.43 \textwidth]{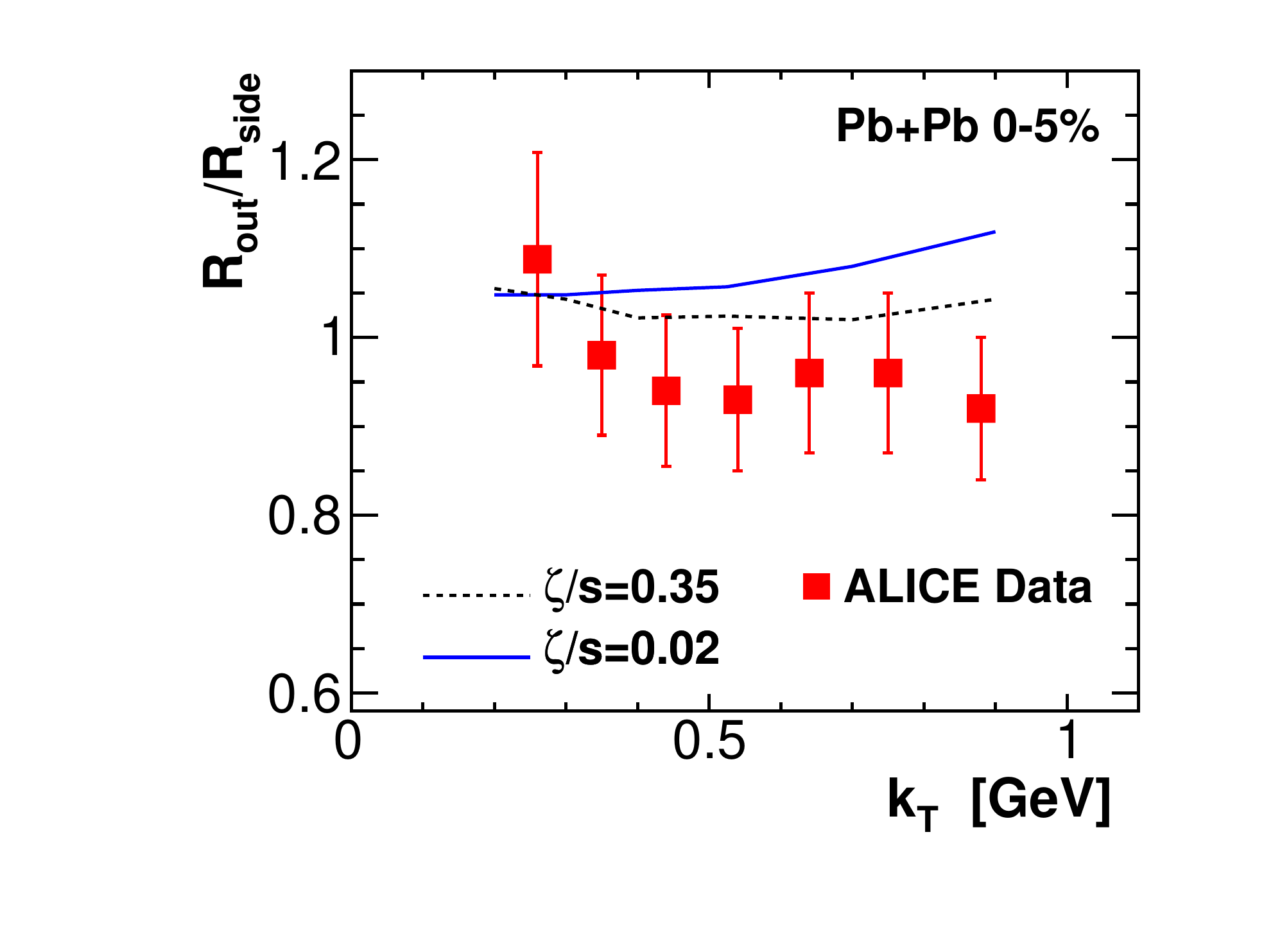}
\vspace{-10mm}
\caption{(color online) The ratio of HBT radii $R_{out}/R_{side}$ for central Pb-Pb collisions at $2760$~GeV. ALICE Collaboration data (squares) are compared to  hydrodynamic calculations using smooth initial conditions
 with (dotted line) and without (solid line) 
a peak in the temperature dependence of bulk viscosity.
\label{fig:ros1}}
\end{figure}

Both calculations, with and without a bulk viscosity peak, reproduce the HBT radii fairly well (Fig. \ref{fig:hbt1}).
The differences in the HBT radii  between the two scenarios are small, less than 5\%. 
The increase of bulk viscosity around the critical temperature does not modify strongly the dynamics.
 Note that the $R_{out}/R_{side}$ ratio decreases when  the bulk viscosity peak is introduced (Fig. \ref{fig:ros1}). 
This decrease may be understood as due to the observed change in the shape of the freeze-out surface \cite{Kisiel:2006is}.
The ratio  $R_{out}/R_{side}$ is larger  when the outside layers of the fireball freeze-out earlier.

\section{Event-by-event hydrodynamic simulations}

\begin{figure}[tb]
\includegraphics[width=0.3 \textwidth]{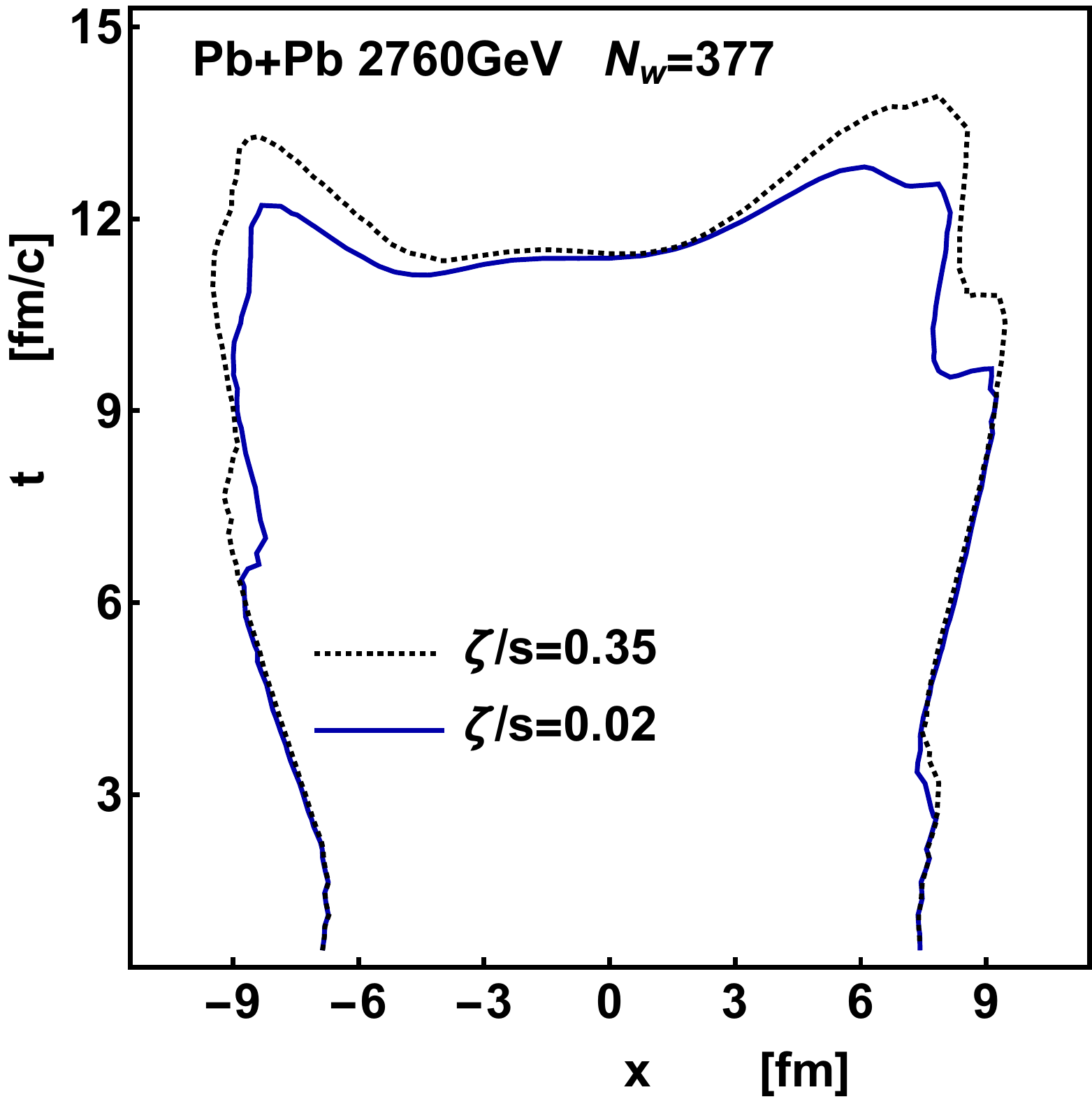}
\caption{(color online) Same as Fig. \ref{fig:fr1} but starting the evolution with one particular initial condition from a Glauber Monte Carlo model.
\label{fig:fr2}}
\end{figure}

\begin{figure}[tb]
\includegraphics[width=0.52 \textwidth]{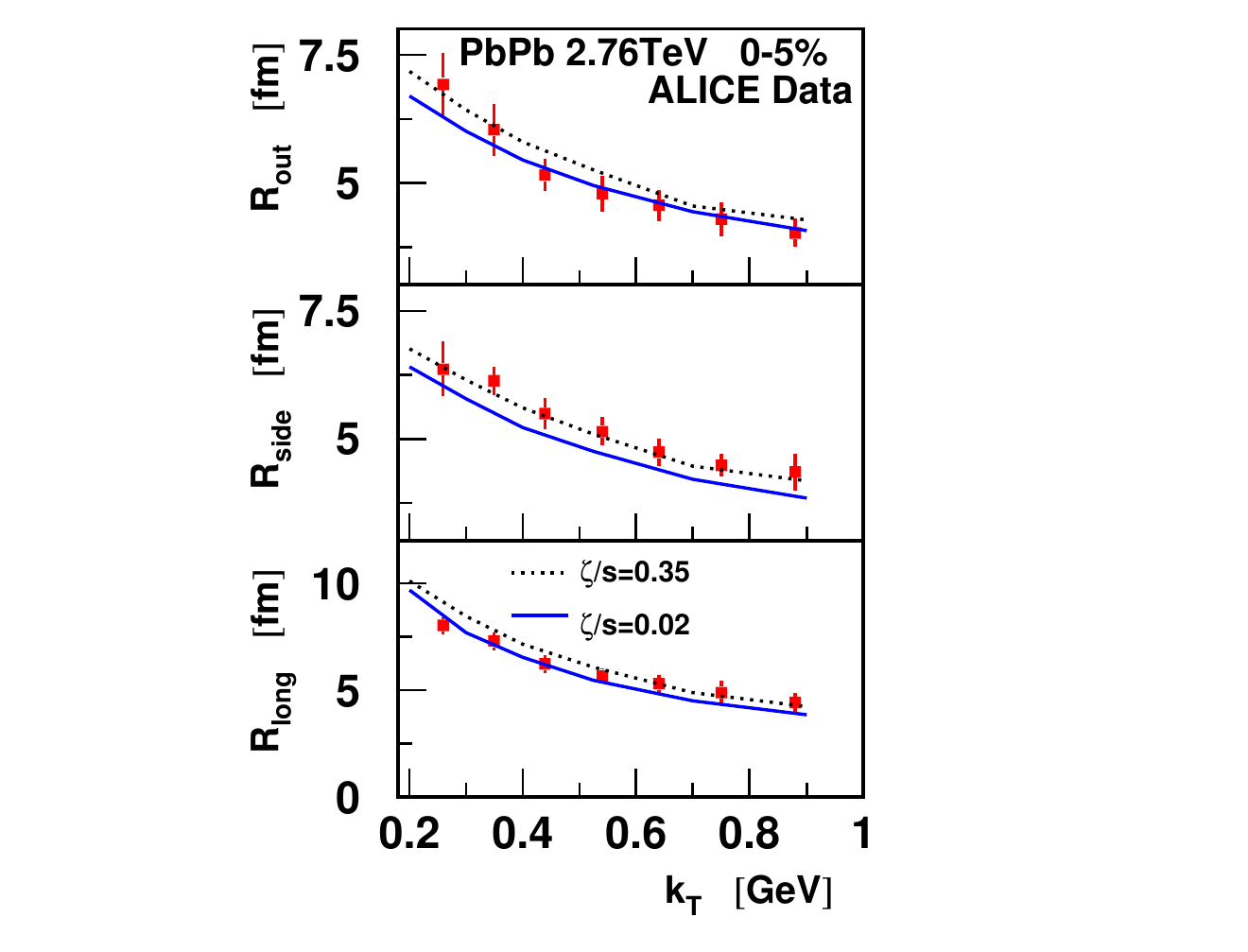}
\caption{(color online) Same as Fig. \ref{fig:hbt1} but for a calculation using an ensemble of fluctuating initial conditions from a Glauber Monte Carlo model.
\label{fig:hbt2}}
\end{figure}

\begin{figure}[tb]
\includegraphics[width=0.43 \textwidth]{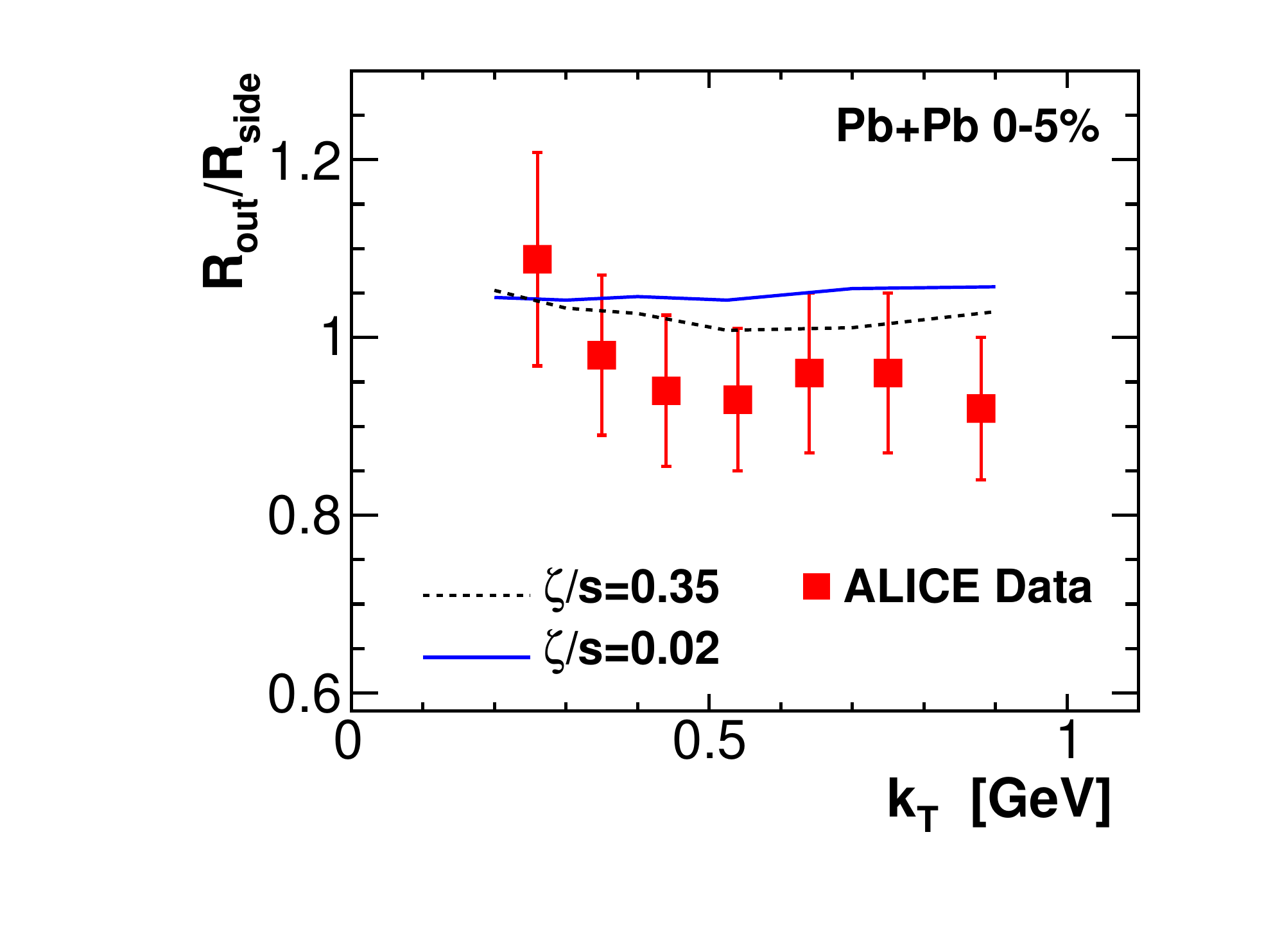}
\vspace{-10mm}
\caption{(color online) Same as Fig. \ref{fig:ros1} but for a calculation using an 
ensemble on fluctuating initial conditions from a Glauber Monte Carlo model.
\label{fig:ros2}}
\end{figure}

A realistic modeling of the collective expansion in heavy-ion collisions requires  hydrodynamic simulations to be performed
 for an ensemble of fluctuating initial conditions. I use a Glauber Monte Carlo model to generate the initial entropy density \cite{Rybczynski:2013yba}. 
The density in the transverse plane is written as a sum of Gaussians of width $0.5$~fm centered at the positions of the
participant nucleons. A contribution of binary collisions is added with $\alpha=0.15$. In Fig. \ref{fig:fr2} are shown  
the freeze-out hypersurfaces corresponding to the same initial Monte Carlo Glauber event (one particular event with 
 377 participant nucleons) but evolved 
with hydrodynamics with or without a peak in the temperature dependence of bulk viscosity. 
 The effect of bulk viscosity is qualitatively similar as for  smooth initial conditions. Quantitatively the difference
 between the two scenarios is larger at the very edge of the fireball, due to larger local gradients of the density in 
event-by-event simulations.
The increase in bulk viscosity hinders the expansion in the outer layers of the fireball, 
where the initial gradients are the largest.  It leads
to a freeze-out hypersurface, where inner layers freeze-out earlier.

For each event a freeze-out hypersurface is defined and the pion pair correlation function is constructed. 
After averaging over events,  HBT radii are extracted from the Gaussian formula (Eq. \ref{eq:pb}). 
For  realistic event-by-event simulations
 one gets similar conclusions as for the expansion from the  smooth initial density discussed in the previous section.
The change in the value of the radii is small (Fig. \ref{fig:hbt2}). The final effect  on the HBT radii is
 to reduce the $R_{out}/R_{side}$ ratio by up to 3\% for the calculation with a bulk viscosity peak (Fig. \ref{fig:ros2}). 
This improves the description of the data.

\section{Conclusions}

I have tested the effect of a peak in the temperature dependence of bulk viscosity  on the size and the life-time of the source
in Pb-Pb collisions at $2760$GeV. The increase of bulk viscosity around the critical temperature slows down the expansion 
of the fireball in the  outer layers, where initially the gradients are the largest and the temperature of 
matter  is close to the critical 
temperature. The relative increase in the life-time of the fireball is small in central Pb-Pb collisions, and limited to the
 outer layers of the fireball.

The effective softening of the equation of state due to the bulk viscosity peak does not lead to an increase of the $R_{out}/R_{side}$ 
ratio. This observation is contrary to expectation from earlier studies of the sensitivity of the HBT radii to the equation
of state \cite{Rischke:1996em,Bozek:2009ty}.
Surprisingly, the ratio $R_{out}/R_{side}$ is reduced when the bulk viscosity peak is introduced.
 Although the expansion is slightly slower with bulk viscosity, the
 shape of the hypersurface is modified in such a way as to reduce 
the problematic $R_{out}/R_{side}$ ratio by a few percent, making it closer to the data.

It would be interesting to include the interferometry radii in the data constraining the parameters of the matter created in
relativistic heavy-ion collisions using exhaustive  Bayesian statistical analyses.
 Existing analysis have studied a matter either without a bulk viscosity peak or they do not take 
the HBT radii in the data used to constraint the parameters
 \cite{Novak:2013bqa,Pratt:2015zsa,Bernhard:2015hxa,Bernhard:2016tnd}. The present result suggests that the presence of a peak in the
temperature dependence of bulk viscosity would be  consistent with an enlarged 
 analysis including HBT radii in the data set. Further insight could be gained by studying the effects of bulk viscosity 
on the HBT radii in small, rapidly expanding systems, as in p+Pb collisions.

\begin{acknowledgments}

Research supported by the Polish Ministry of Science and Higher Education (MNiSW), by the National
Science Centre grant DEC-2012/06/A/ST2/00390, as well as by PL-Grid Infrastructure. 

\end{acknowledgments}

\bibliography{../hydr}

\begin{thebibliography}{10}%
\makeatletter
\providecommand \@ifxundefined [1]{%
 \ifx #1\undefined \expandafter \@firstoftwo
 \else \expandafter \@secondoftwo
\fi
}%
\providecommand \@ifnum [1]{%
 \ifnum #1\expandafter \@firstoftwo
 \else \expandafter \@secondoftwo
\fi
}%
\providecommand \enquote [1]{``#1''}%
\providecommand \bibnamefont  [1]{#1}%
\providecommand \bibfnamefont [1]{#1}%
\providecommand \citenamefont [1]{#1}%
\providecommand\href[0]{\@sanitize\@href}%
\providecommand\@href[1]{\endgroup\@@startlink{#1}\endgroup\@@href}%
\providecommand\@@href[1]{#1\@@endlink}%
\providecommand \@sanitize [0]{\begingroup\catcode`\&12\catcode`\#12\relax}%
\@ifxundefined \pdfoutput {\@firstoftwo}{%
 \@ifnum{\z@=\pdfoutput}{\@firstoftwo}{\@secondoftwo}%
}{%
 \providecommand\@@startlink[1]{\leavevmode\special{html:<a href="#1">}}%
 \providecommand\@@endlink[0]{\special{html:</a>}}%
}{%
 \providecommand\@@startlink[1]{%
  \leavevmode
  \pdfstartlink
   attr{/Border[0 0 1 ]/H/I/C[0 1 1]}%
   user{/Subtype/Link/A<</Type/Action/S/URI/URI(#1)>>}%
  \relax
 }%
 \providecommand\@@endlink[0]{\pdfendlink}%
}%
\providecommand \url  [0]{\begingroup\@sanitize \@url }%
\providecommand \@url [1]{\endgroup\@href {#1}{\urlprefix}}%
\providecommand \urlprefix [0]{URL }%
\providecommand \Eprint[0]{\href }%
\@ifxundefined \urlstyle {%
  \providecommand \doi [1]{doi:\discretionary{}{}{}#1}%
}{%
  \providecommand \doi [0]{doi:\discretionary{}{}{}\begingroup
  \urlstyle{rm}\Url }%
}%
\providecommand \doibase [0]{http://dx.doi.org/}%
\providecommand \Doi[1]{\href{\doibase#1}}%
\providecommand \bibAnnote [3]{%
  \BibitemShut{#1}%
  \begin{quotation}\noindent
    \textsc{Key:}\ #2\\\textsc{Annotation:}\ #3%
  \end{quotation}%
}%
\providecommand \bibAnnoteFile [2]{%
  \IfFileExists{#2}{\bibAnnote {#1} {#2} {\input{#2}}}{}%
}%
\providecommand \typeout [0]{\immediate \write \m@ne }%
\providecommand \selectlanguage [0]{\@gobble}%
\providecommand \bibinfo [0]{\@secondoftwo}%
\providecommand \bibfield [0]{\@secondoftwo}%
\providecommand \translation [1]{[#1]}%
\providecommand \BibitemOpen[0]{}%
\providecommand \bibitemStop [0]{}%
\providecommand \bibitemNoStop [0]{.\EOS\space}%
\providecommand \EOS [0]{\spacefactor3000\relax}%
\providecommand \BibitemShut [1]{\csname bibitem#1\endcsname}%
\bibitem{Lisa:2005dd}%
  \BibitemOpen
  \bibfield{author}{%
  \bibinfo {author} {\bibfnamefont{M.~A.}\ \bibnamefont{Lisa}}, \bibinfo
  {author} {\bibfnamefont{S.}~\bibnamefont{Pratt}}, \bibinfo {author}
  {\bibfnamefont{R.}~\bibnamefont{Soltz}},\ and\ \bibinfo {author}
  {\bibfnamefont{U.}~\bibnamefont{Wiedemann}},\ }%
  \bibfield{journal}{%
  \bibinfo {journal} {Ann. Rev. Nucl. Part. Sci.}\ }%
  \textbf{\bibinfo {volume} {55}},\ \bibinfo {pages} {357} (\bibinfo {year}
  {2005})%
  \bibAnnoteFile{NoStop}{Lisa:2005dd}%
\bibitem{Wiedemann:1999qn}%
  \BibitemOpen
  \bibfield{author}{%
  \bibinfo {author} {\bibfnamefont{U.~A.}\ \bibnamefont{Wiedemann}}\ and\
  \bibinfo {author} {\bibfnamefont{U.~W.}\ \bibnamefont{Heinz}},\ }%
  \bibfield{journal}{%
  \Doi{10.1016/S0370-1573(99)00032-0}{\bibinfo {journal} {Phys. Rept.}}\ }%
  \textbf{\bibinfo {volume} {319}},\ \bibinfo {pages} {145} (\bibinfo {year}
  {1999})%
  \bibAnnoteFile{NoStop}{Wiedemann:1999qn}%
\bibitem{Broniowski:2008vp}%
  \BibitemOpen
  \bibfield{author}{%
  \bibinfo {author} {\bibfnamefont{W.}~\bibnamefont{Broniowski}}, \bibinfo
  {author} {\bibfnamefont{M.}~\bibnamefont{Chojnacki}}, \bibinfo {author}
  {\bibfnamefont{W.}~\bibnamefont{Florkowski}},\ and\ \bibinfo {author}
  {\bibfnamefont{A.}~\bibnamefont{Kisiel}},\ }%
  \bibfield{journal}{%
  \Doi{10.1103/PhysRevLett.101.022301}{\bibinfo {journal} {Phys. Rev. Lett.}}\
  }%
  \textbf{\bibinfo {volume} {101}},\ \bibinfo {pages} {022301} (\bibinfo {year}
  {2008})%
  \bibAnnoteFile{NoStop}{Broniowski:2008vp}%
\bibitem{Pratt:2008qv}%
  \BibitemOpen
  \bibfield{author}{%
  \bibinfo {author} {\bibfnamefont{S.}~\bibnamefont{Pratt}},\ }%
  \bibfield{journal}{%
  \Doi{10.1103/PhysRevLett.102.232301}{\bibinfo {journal} {Phys. Rev. Lett.}}\
  }%
  \textbf{\bibinfo {volume} {102}},\ \bibinfo {pages} {232301} (\bibinfo {year}
  {2009})%
  \bibAnnoteFile{NoStop}{Pratt:2008qv}%
\bibitem{Bertsch:1989vn}%
  \BibitemOpen
  \bibfield{author}{%
  \bibinfo {author} {\bibfnamefont{G.~F.}\ \bibnamefont{Bertsch}},\ }%
  \bibfield{journal}{%
  \bibinfo {journal} {Nucl. Phys.}\ }%
  \textbf{\bibinfo {volume} {A498}},\ \bibinfo {pages} {173c} (\bibinfo {year}
  {1989})%
  \bibAnnoteFile{NoStop}{Bertsch:1989vn}%
\bibitem{Pratt:1986cc}%
  \BibitemOpen
  \bibfield{author}{%
  \bibinfo {author} {\bibfnamefont{S.}~\bibnamefont{Pratt}},\ }%
  \bibfield{journal}{%
  \Doi{10.1103/PhysRevD.33.1314}{\bibinfo {journal} {Phys. Rev.}}\ }%
  \textbf{\bibinfo {volume} {D33}},\ \bibinfo {pages} {1314} (\bibinfo {year}
  {1986})%
  \bibAnnoteFile{NoStop}{Pratt:1986cc}%
\bibitem{Rischke:1996em}%
  \BibitemOpen
  \bibfield{author}{%
  \bibinfo {author} {\bibfnamefont{D.~H.}\ \bibnamefont{Rischke}}\ and\
  \bibinfo {author} {\bibfnamefont{M.}~\bibnamefont{Gyulassy}},\ }%
  \bibfield{journal}{%
  \bibinfo {journal} {Nucl. Phys.}\ }%
  \textbf{\bibinfo {volume} {A608}},\ \bibinfo {pages} {479} (\bibinfo {year}
  {1996})%
  \bibAnnoteFile{NoStop}{Rischke:1996em}%
\bibitem{Bozek:2009ty}%
  \BibitemOpen
  \bibfield{author}{%
  \bibinfo {author} {\bibfnamefont{P.}~\bibnamefont{Bo\.zek}}\ and\ \bibinfo
  {author} {\bibfnamefont{I.}~\bibnamefont{Wyskiel}},\ }%
  \bibfield{journal}{%
  \Doi{10.1103/PhysRevC.79.044916}{\bibinfo {journal} {Phys. Rev.}}\ }%
  \textbf{\bibinfo {volume} {C79}},\ \bibinfo {pages} {044916} (\bibinfo {year}
  {2009})%
  \bibAnnoteFile{NoStop}{Bozek:2009ty}%
\bibitem{Bozek:2010er}%
  \BibitemOpen
  \bibfield{author}{%
  \bibinfo {author} {\bibfnamefont{P.}~\bibnamefont{Bo\.zek}},\ }%
  \bibfield{journal}{%
  \Doi{10.1103/PhysRevC.83.044910}{\bibinfo {journal} {Phys. Rev.}}\ }%
  \textbf{\bibinfo {volume} {C83}},\ \bibinfo {pages} {044910} (\bibinfo {year}
  {2011})%
  \bibAnnoteFile{NoStop}{Bozek:2010er}%
\bibitem{Karpenko:2012yf}%
  \BibitemOpen
  \bibfield{author}{%
  \bibinfo {author} {\bibfnamefont{I.~A.}\ \bibnamefont{Karpenko}}, \bibinfo
  {author} {\bibfnamefont{Y.~M.}\ \bibnamefont{Sinyukov}},\ and\ \bibinfo
  {author} {\bibfnamefont{K.}~\bibnamefont{Werner}},\ }%
  \bibfield{journal}{%
  \Doi{10.1103/PhysRevC.87.024914}{\bibinfo {journal} {Phys. Rev.}}\ }%
  \textbf{\bibinfo {volume} {C87}},\ \bibinfo {pages} {024914} (\bibinfo {year}
  {2013})%
  \bibAnnoteFile{NoStop}{Karpenko:2012yf}%
\bibitem{Karsch:2007jc}%
  \BibitemOpen
  \bibfield{author}{%
  \bibinfo {author} {\bibfnamefont{F.}~\bibnamefont{Karsch}}, \bibinfo {author}
  {\bibfnamefont{D.}~\bibnamefont{Kharzeev}},\ and\ \bibinfo {author}
  {\bibfnamefont{K.}~\bibnamefont{Tuchin}},\ }%
  \bibfield{journal}{%
  \Doi{10.1016/j.physletb.2008.01.080}{\bibinfo {journal} {Phys. Lett.}}\ }%
  \textbf{\bibinfo {volume} {B663}},\ \bibinfo {pages} {217} (\bibinfo {year}
  {2008})%
  \bibAnnoteFile{NoStop}{Karsch:2007jc}%
\bibitem{Denicol:2009am}%
  \BibitemOpen
  \bibfield{author}{%
  \bibinfo {author} {\bibfnamefont{G.~S.}\ \bibnamefont{Denicol}}, \bibinfo
  {author} {\bibfnamefont{T.}~\bibnamefont{Kodama}}, \bibinfo {author}
  {\bibfnamefont{T.}~\bibnamefont{Koide}},\ and\ \bibinfo {author}
  {\bibfnamefont{P.}~\bibnamefont{Mota}},\ }%
  \bibfield{journal}{%
  \Doi{10.1103/PhysRevC.80.064901}{\bibinfo {journal} {Phys. Rev.}}\ }%
  \textbf{\bibinfo {volume} {C80}},\ \bibinfo {pages} {064901} (\bibinfo {year}
  {2009})%
  \bibAnnoteFile{NoStop}{Denicol:2009am}%
\bibitem{Song:2009rh}%
  \BibitemOpen
  \bibfield{author}{%
  \bibinfo {author} {\bibfnamefont{H.}~\bibnamefont{Song}}\ and\ \bibinfo
  {author} {\bibfnamefont{U.~W.}\ \bibnamefont{Heinz}},\ }%
  \bibfield{journal}{%
  \Doi{10.1103/PhysRevC.81.024905}{\bibinfo {journal} {Phys. Rev.}}\ }%
  \textbf{\bibinfo {volume} {{C81}}},\ \bibinfo {pages} {{024905}} (\bibinfo
  {year} {2010})%
  \bibAnnoteFile{NoStop}{Song:2009rh}%
\bibitem{Ryu:2015vwa}%
  \BibitemOpen
  \bibfield{author}{%
  \bibinfo {author} {\bibfnamefont{S.}~\bibnamefont{Ryu}}, \bibinfo {author}
  {\bibfnamefont{J.~F.}\ \bibnamefont{Paquet}}, \bibinfo {author}
  {\bibfnamefont{C.}~\bibnamefont{Shen}}, \bibinfo {author}
  {\bibfnamefont{G.~S.}\ \bibnamefont{Denicol}}, \bibinfo {author}
  {\bibfnamefont{B.}~\bibnamefont{Schenke}}, \bibinfo {author}
  {\bibfnamefont{S.}~\bibnamefont{Jeon}},\ and\ \bibinfo {author}
  {\bibfnamefont{C.}~\bibnamefont{Gale}},\ }%
  \bibfield{journal}{%
  \Doi{10.1103/PhysRevLett.115.132301}{\bibinfo {journal} {Phys. Rev. Lett.}}\
  }%
  \textbf{\bibinfo {volume} {115}},\ \bibinfo {pages} {132301} (\bibinfo {year}
  {2015})%
  \bibAnnoteFile{NoStop}{Ryu:2015vwa}%
\bibitem{Bernhard:2016tnd}%
  \BibitemOpen
  \bibfield{author}{%
  \bibinfo {author} {\bibfnamefont{J.~E.}\ \bibnamefont{Bernhard}}, \bibinfo
  {author} {\bibfnamefont{J.~S.}\ \bibnamefont{Moreland}}, \bibinfo {author}
  {\bibfnamefont{S.~A.}\ \bibnamefont{Bass}}, \bibinfo {author}
  {\bibfnamefont{J.}~\bibnamefont{Liu}},\ and\ \bibinfo {author}
  {\bibfnamefont{U.}~\bibnamefont{Heinz}},\ }%
  \bibfield{journal}{%
  \Doi{10.1103/PhysRevC.94.024907}{\bibinfo {journal} {Phys. Rev.}}\ }%
  \textbf{\bibinfo {volume} {C94}},\ \bibinfo {pages} {024907} (\bibinfo {year}
  {2016})%
  \bibAnnoteFile{NoStop}{Bernhard:2016tnd}%
\bibitem{Schenke:2010rr}%
  \BibitemOpen
  \bibfield{author}{%
  \bibinfo {author} {\bibfnamefont{B.}~\bibnamefont{Schenke}}, \bibinfo
  {author} {\bibfnamefont{S.}~\bibnamefont{Jeon}},\ and\ \bibinfo {author}
  {\bibfnamefont{C.}~\bibnamefont{Gale}},\ }%
  \bibfield{journal}{%
  \Doi{10.1103/PhysRevLett.106.042301}{\bibinfo {journal} {Phys. Rev. Lett.}}\
  }%
  \textbf{\bibinfo {volume} {106}},\ \bibinfo {pages} {042301} (\bibinfo {year}
  {2011})%
  \bibAnnoteFile{NoStop}{Schenke:2010rr}%
\bibitem{Bozek:2009dw}%
  \BibitemOpen
  \bibfield{author}{%
  \bibinfo {author} {\bibfnamefont{P.}~\bibnamefont{Bo\.zek}},\ }%
  \bibfield{journal}{%
  \bibinfo {journal} {Phys. Rev.}\ }%
  \textbf{\bibinfo {volume} {C81}},\ \bibinfo {pages} {034909} (\bibinfo {year}
  {2010})%
  \bibAnnoteFile{NoStop}{Bozek:2009dw}%
\bibitem{Chojnacki:2011hb}%
  \BibitemOpen
  \bibfield{author}{%
  \bibinfo {author} {\bibfnamefont{M.}~\bibnamefont{Chojnacki}}, \bibinfo
  {author} {\bibfnamefont{A.}~\bibnamefont{Kisiel}}, \bibinfo {author}
  {\bibfnamefont{W.}~\bibnamefont{Florkowski}},\ and\ \bibinfo {author}
  {\bibfnamefont{W.}~\bibnamefont{Broniowski}},\ }%
  \bibfield{journal}{%
  \Doi{10.1016/j.cpc.2011.11.018}{\bibinfo {journal} {Comput. Phys. Commun.}}\
  }%
  \textbf{\bibinfo {volume} {183}},\ \bibinfo {pages} {746} (\bibinfo {year}
  {2012})%
  \bibAnnoteFile{NoStop}{Chojnacki:2011hb}%
\bibitem{Bozek:2011ua}%
  \BibitemOpen
  \bibfield{author}{%
  \bibinfo {author} {\bibfnamefont{P.}~\bibnamefont{Bo\.zek}},\ }%
  \bibfield{journal}{%
  \Doi{10.1103/PhysRevC.85.034901}{\bibinfo {journal} {Phys. Rev.}}\ }%
  \textbf{\bibinfo {volume} {C85}},\ \bibinfo {pages} {034901} (\bibinfo {year}
  {2012})%
  \bibAnnoteFile{NoStop}{Bozek:2011ua}%
\bibitem{Kisiel:2006is}%
  \BibitemOpen
  \bibfield{author}{%
  \bibinfo {author} {\bibfnamefont{A.}~\bibnamefont{Kisiel}}, \bibinfo {author}
  {\bibfnamefont{W.}~\bibnamefont{Florkowski}}, \bibinfo {author}
  {\bibfnamefont{W.}~\bibnamefont{Broniowski}},\ and\ \bibinfo {author}
  {\bibfnamefont{J.}~\bibnamefont{Pluta}},\ }%
  \bibfield{journal}{%
  \Doi{10.1103/PhysRevC.73.064902}{\bibinfo {journal} {Phys. Rev.}}\ }%
  \textbf{\bibinfo {volume} {C73}},\ \bibinfo {pages} {064902} (\bibinfo {year}
  {2006})%
  \bibAnnoteFile{NoStop}{Kisiel:2006is}%
\bibitem{Rybczynski:2013yba}%
  \BibitemOpen
  \bibfield{author}{%
  \bibinfo {author} {\bibfnamefont{M.}~\bibnamefont{Rybczy\'nski}}, \bibinfo
  {author} {\bibfnamefont{G.}~\bibnamefont{Stefanek}}, \bibinfo {author}
  {\bibfnamefont{W.}~\bibnamefont{Broniowski}},\ and\ \bibinfo {author}
  {\bibfnamefont{P.}~\bibnamefont{Bo\.zek}},\ }%
  \bibfield{journal}{%
  \Doi{10.1016/j.cpc.2014.02.016}{\bibinfo {journal} {Comput. Phys. Commun.}}\
  }%
  \textbf{\bibinfo {volume} {185}},\ \bibinfo {pages} {1759} (\bibinfo {year}
  {2014})%
  \bibAnnoteFile{NoStop}{Rybczynski:2013yba}%
\bibitem{Novak:2013bqa}%
  \BibitemOpen
  \bibfield{author}{%
  \bibinfo {author} {\bibfnamefont{J.}~\bibnamefont{Novak}}, \bibinfo {author}
  {\bibfnamefont{K.}~\bibnamefont{Novak}}, \bibinfo {author}
  {\bibfnamefont{S.}~\bibnamefont{Pratt}}, \bibinfo {author}
  {\bibfnamefont{J.}~\bibnamefont{Vredevoogd}}, \bibinfo {author}
  {\bibfnamefont{C.}~\bibnamefont{Coleman-Smith}},\ and\ \bibinfo {author}
  {\bibfnamefont{R.}~\bibnamefont{Wolpert}},\ }%
  \bibfield{journal}{%
  \Doi{10.1103/PhysRevC.89.034917}{\bibinfo {journal} {Phys. Rev.}}\ }%
  \textbf{\bibinfo {volume} {C89}},\ \bibinfo {pages} {034917} (\bibinfo {year}
  {2014})%
  \bibAnnoteFile{NoStop}{Novak:2013bqa}%
\bibitem{Pratt:2015zsa}%
  \BibitemOpen
  \bibfield{author}{%
  \bibinfo {author} {\bibfnamefont{S.}~\bibnamefont{Pratt}}, \bibinfo {author}
  {\bibfnamefont{E.}~\bibnamefont{Sangaline}}, \bibinfo {author}
  {\bibfnamefont{P.}~\bibnamefont{Sorensen}},\ and\ \bibinfo {author}
  {\bibfnamefont{H.}~\bibnamefont{Wang}},\ }%
  \bibfield{journal}{%
  \Doi{10.1103/PhysRevLett.114.202301}{\bibinfo {journal} {Phys. Rev. Lett.}}\
  }%
  \textbf{\bibinfo {volume} {114}},\ \bibinfo {pages} {202301} (\bibinfo {year}
  {2015})%
  \bibAnnoteFile{NoStop}{Pratt:2015zsa}%
\bibitem{Bernhard:2015hxa}%
  \BibitemOpen
  \bibfield{author}{%
  \bibinfo {author} {\bibfnamefont{J.~E.}\ \bibnamefont{Bernhard}}, \bibinfo
  {author} {\bibfnamefont{P.~W.}\ \bibnamefont{Marcy}}, \bibinfo {author}
  {\bibfnamefont{C.~E.}\ \bibnamefont{Coleman-Smith}}, \bibinfo {author}
  {\bibfnamefont{S.}~\bibnamefont{Huzurbazar}}, \bibinfo {author}
  {\bibfnamefont{R.~L.}\ \bibnamefont{Wolpert}},\ and\ \bibinfo {author}
  {\bibfnamefont{S.~A.}\ \bibnamefont{Bass}},\ }%
  \bibfield{journal}{%
  \Doi{10.1103/PhysRevC.91.054910}{\bibinfo {journal} {Phys. Rev.}}\ }%
  \textbf{\bibinfo {volume} {C91}},\ \bibinfo {pages} {054910} (\bibinfo {year}
  {2015})%
  \bibAnnoteFile{NoStop}{Bernhard:2015hxa}%
\end{thebibliography}%

\end{document}